\documentstyle[aps,epsfig,a4]{revtex}
\begin{document}
\pagestyle{myheadings}\markright{$J/\Psi$ suppression in colliding nuclei: statistical model analysis}%
\draft%

\title{$J/\Psi$ suppression in colliding nuclei: statistical model analysis}
\author{Dariusz Prorok and Ludwik Turko}
\address{Institute of Theoretical Physics, University of
Wroc{\l}aw,\\ Pl. Maksa Borna 9, 50-204  Wroc{\l}aw, Poland}
\date{December 27, 2000}
\maketitle
\begin{abstract}
We consider the $J/\Psi$ suppression at a high energy heavy ion
collision. An ideal gas of massive hadrons in thermal and chemical
equilibrium is formed in the central region. The finite-size gas
expands longitudinally in accordance with Bjorken law. The
transverse expansion in a form of the rarefaction wave is taken
into account. We show that $J/\Psi$ suppression in such an
environment, when combined with the disintegration in nuclear
matter, gives correct evaluation of NA38 and NA50 data in a broad
range of initial energy densities.
\end{abstract}
\section {Introduction }

Since the paper of Matsui and Satz \cite{Matsui:1986dk} there is a
steady interest in the problem of $J/\Psi$ suppression in a
heavy-ion collision. The question is if this suppression can be
treated as a signature for a quark-gluon plasma (QGP) or if it can
be explained by $J/\Psi$ absorption in a hadron gas which appears
instead of the QGP in the central rapidity region (CRR) of the
collision
\cite{Ftacnik:1988qv,Vogt:1988fj,Gavin:1988hs,Blaizot:1989ec,Vogt:1999cu}.
In the following paper we shall explore the latter possibility.

The existence of the QGP --- the novel phase of hadron matter, has
been predicted upon lattice QCD calculations (for a review see
\cite{Karsch:2001vs;2000vy} and references therein). One gets from
those calculations the critical temperature $T_{c}$ in the range
of 150 - 270 MeV. The upper limit belongs to a pure $SU(3)$
theory, whereas adding quarks causes lowering of $T_{c}$ even to
about 150 MeV ($N_{f}\geq 3$). All lattice calculations have been
performed for zero quark and baryon chemical potential only.
Additionally to the fact that lattice estimations could not be
treated literally, in relativistic heavy ion collisions there is a
finite baryon chemical potential. As a result, the above-mentioned
evaluations of $T_{c}$ can be understood only as a measure of the
order of the magnitude of real $T_{c}$. Therefore, the assumption
that at initial temperatures around 200 MeV (as in our model) a
hadron gas still exists is realistic entirely. As far as the
hadron gas is considered itself in that range of temperature, it
was shown in \cite{Bratkovskaya:2000qy} that the temperature
increases with energy density when continuum excitations (string
degrees of freedom) are not taken into account. Continuum
excitations gives the limiting temperature. This reproduces
results of the Hagedorn bootstrap
model\cite{Hagedorn:1965st;1972gv}. The thermodynamical analysis
can also be based on particle ratios \cite{Stachel:1999rc} and
this gives the temperature as in our model. The same result is
obtained from the microscopic cascade model (see e.g.
\cite{Bravina:1999dh}).

In the following paper we shall continue our previous
investigations \cite{Prorok:1994fb} into the problem of $J/\Psi$
suppression observed in a heavy-ion collision (for experimental
data see e.g. \cite{Abr,Abr50} and references therein). Now, we
shall focus on the dependence of the suppression on the initial
energy density reached in the CRR.

In our model, $J/\Psi$ suppression is the result of a $c\bar c$
state absorption in a dense hadronic matter through interactions
of the type

\begin{equation}
c\bar c+h \longrightarrow D+\bar D+X\\ ,  \label{psiabs}
\end{equation}


\noindent where $h$ denotes a hadron, $D$ is a charm meson and $X$
means a particle which is necessary to conserve the charge, baryon
number or strangeness.

First, the $J/\Psi$ is absorbed by nucleons in colliding nuclei
\cite{Gerschel:1988wn}. Later, the $J/\Psi$ is absorbed by
secondary hadrons produced in the collision. In the simpler
scenario those secondary hadrons expand longitudinally along
collision axis \cite{Gavin:1990gm,Brodsky:1988xz} and the time
evolution is given by Bjorken's scaling dynamics
\cite{Bjorken:1983qr}. The importance of secondaries scattering
becomes more and more important at higher energies since secondary
production grows with energy as well. We assume that at very high
energies secondaries form a dense hadronic gas in a state of
thermodynamical equilibrium. A chemistry will be given by equation
of state of an ideal gas consisting of different species of
massive hadrons.

A problem of thermal and chemical equilibration of an hadronic
fireball formatted in a heavy collision is still far from being
solved. Recent results
\cite{Bravina:1999dh,Bratkovskaya:2000qy,Bellwied:2000mi}, based
on parton cascade models, suggest that in big systems the
equilibration time for heavier particles is longer than for
lighter particles. The thermal equilibrium is established quickly,
in about 5 fm/c, much faster than the chemical equilibrium.

To keep our model as simple as possible we consider the one
temperature model. This allows us to reduce the number of external
parameters. It is obvious that any new parameter would result in a
better fit to experimental data.

The hadronic matter is in a state of an ideal gas of massive
hadrons in thermal and chemical equilibrium and consists of all
species up to $\Omega^{-}$ baryon. Time evolution is given here by
conservation laws combined with assumptions about the space-time
structure of the system. A corresponding equation of state of the
ideal gas makes then possible to express gas parameters such as
temperature and chemical potentials as functions of time.

\begin{figure}
\begin{center}{
{\epsfig{file=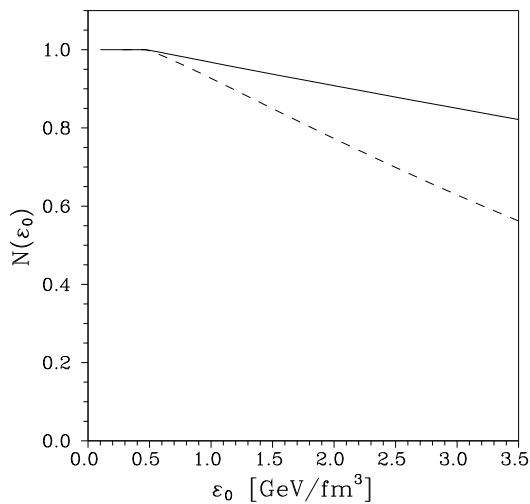,width=7cm}} }\end{center}
\caption{Comparison of suppression of the pure $J/\Psi$'s in the
cooling hadron gas for two values of the power $a$ in
approximation $T(t)\protect\cong T_0 \cdot t^{-a}$: $a= {1\over
3}$ (solid) and $a = {1\over 5.6}$ (dashed).}
\label{Fig.1.}
\end{figure}

An ideal gas of real hadrons has a very interesting feature: it
cools much slower than a pion gas when expands longitudinally. We
have checked numerically that for the initial energy densities
$\epsilon_0$ corresponding to initial temperatures $T_{0}$ of the
order of 200 MeV and for the freeze-out $T_{f.o.}$ not lower than
about 100 MeV, the time dependence of the temperature of the
expanding gas still keeps the well-known form $T(t)=T_{0} \cdot
t^{-a}$ (we put $t_{0}=1$ fm). Only the exponent $a$ changes from
${1 \over 3}$ for massless pions to the values ${1 \over
{5.6}}$-${1 \over {5.3}}$ for massive realistic hadrons. As a
result, the time of the freeze-out $t_{f.o}$ is much greater for
the hadron gas than for the pion one. For instance, when we take
$T_{0}=200$ MeV and $T_{f.o.}=140$ MeV we obtain $t_{f.o.}={7.37}$
fm ($a={1 \over {5.6}}$) for the hadron gas and $t_{f.o.}={2.9}$
fm ($a={1 \over 3}$) for the pion gas. The lower $T_{f.o.}$, the
stronger difference. For $T_{0}=200$ MeV and $T_{f.o.}=100$ MeV we
have $t_{f.o.}={48.5}$ fm ($a={1 \over {5.6}}$) for the hadron gas
and $t_{f.o.}={8}$ fm ($a={1 \over 3}$) for the pion gas. This has
a direct consequence for $J/\Psi$ suppression: the longer the
system lasts, the deeper suppression causes (see
Fig.\,\ref{Fig.1.}).

We are going to calculate a survival factor for $J/\Psi$ when new,
more realistic conditions are taken into account. We consider a
hadronic gas which is produced in the CRR region. This gas expands
both longitudinally and transversely. The longitudinal expansion
is a traditional adiabatic hydrodynamical evolution
\cite{Bjorken:1983qr}, the transverse expansion is considered as
the rarefaction wave. An initial energy density depends now on the
impact parameter $b$ and on the geometry of the collision.

The concept of the description of $J/\Psi$ absorption in the hot
hadronic matter is based on the application of the relativistic
kinetic equation as it was postulated in \cite{Blaizot:1989ec}. As
the hadronic medium a pion gas was taken into account there. Our
hadronic medium consists of hadrons from the lowest up to
$\Omega^{-}$ as constituents of the matter in the CRR.  This gives
different physical properties important for the absorption
processes as compared to the absorption in a pion gas --- cooling
time, threshold effects, etc. It allows us to consider here also a
case with non-zero baryon number density. We take into account
here some physical effects which were neglected or not fully
treated in \cite{Blaizot:1989ec}. Thresholds for $J/\Psi-hadron$
reactions as well as relative velocities are examined completely,
but both these effects are ignored in final calculations in
\cite{Blaizot:1989ec}. In addition to the CRR also $J/\Psi$
absorption in the nuclear matter is considered simultaneously
here. In the final estimations of $J/\Psi$ survival fraction, the
Woods-Saxon nuclear matter density distribution is used in the
presented model.

As far as the hydrodynamic is concerned, the full solution of
hydrodynamic equations (for the baryon number density equal to
zero) with the use of the method developed in \cite{Baym:1983sr}
was explored in \cite{Blaizot:1989ec}, but with different initial
temperature profile. Here, for simplicity of numerical
calculations we assumed the uniform initial temperature
distribution with the sharp edge at the border established by
nuclei radii (see Fig.\,\ref{Fig.2.}). For such an initial
distribution, and for a central collision and the baryon number
density equal to zero, hydrodynamic equations are solved
numerically in \cite{Baym:1983sr}. The resulting evolution can be
decomposed into the longitudinal expansion inside a slice bordered
by the front of the rarefaction wave and the transverse expansion
which is superimposed outside of the wave. But the temperature
decreases very quickly outside the wave (at least at times not
greater than the half of the freeze-out time), so it is reasonable
to ignore the influence of the flow outside of the rarefaction
wave as we do in our model. Since we are dealing with small but
nevertheless non-zero baryon number densities, we assume that the
full hydrodynamic evolution looks like the same in this case. For
the longitudinal expansion --- the most important part of the
hydrodynamical evolution, this assumption is confirmed by our
investigations into the time dependence of the temperature of the
hadron gas, where the same pattern (see (\ref{temperature})) has
been obtained approximately as in the case of a massless
baryonless ideal gas \cite{Prorok:1995xz,Prorok:2001ut}.

The dependence of $J/\Psi$ suppression on the impact parameter $b$
is treated here more rigorously than in \cite{Blaizot:1989ec}. In
fact the only approximation of the survival factor (with respect
to its exact expression for the case of the finite-size effects
and the transverse expansion taken into account) is done by
treating the time $J/\Psi$ needs to escape from the hadron matter
in the transverse plane on the average (over the initial $J/\Psi$
positions). The effective freeze-out time $t_{f}$ (denoted by
$t_{final}$ herein) is therefore the function of both the impact
parameter and $J/\Psi$ transverse momentum, whereas $t_{f}$ is
evaluated only for given $b$ (for some assumed average $J/\Psi$
transverse momentum, namely equal to 1 GeV) in
\cite{Blaizot:1989ec}.

\section { The expanding hadron gas }
\label{hadgas}

For an ideal hadron gas in thermal and chemical equilibrium, which
consists of $l$ species of particles, energy density $\epsilon$,
baryon number density $n_{B}$, strangeness density $n_{S}$ and
entropy density $s$ read ($\hbar=c=1$ always)

\begin{mathletters}
\label{eqstate}
\begin{equation}
\epsilon = { 1 \over {2\pi^{2}}} \sum_{i=1}^{l} (2s_{i}+1)
\int_{0}^{\infty}dp\,{ { p^{2}E_{i} } \over { \exp \left\{ {{
E_{i} - \mu_{i} } \over T} \right\} + g_{i} } } \ , \label{energy}
\end{equation}

\begin{equation}
n_{B}={ 1 \over {2\pi^{2}}} \sum_{i=1}^{l} (2s_{i}+1)
\int_{0}^{\infty}dp\,{ { p^{2}B_{i} } \over { \exp \left\{ {{
E_{i} - \mu_{i} } \over T} \right\} + g_{i} } } \ ,
\label{barnumb}
\end{equation}

\begin{equation}
n_{S}={1 \over {2\pi^{2}}} \sum_{i=1}^{l} (2s_{i}+1)
\int_{0}^{\infty}dp\,{ { p^{2}S_{i} } \over { \exp \left\{ {{
E_{i} - \mu_{i} } \over T} \right\} + g_{i} } } \ ,
\label{strange}
\end{equation}

\begin{equation}
s={1 \over {6\pi^{2}T^{2}} } \sum_{i=1}^{l} (2s_{i}+1)
\int_{0}^{\infty}dp\, { {p^{4}} \over { E_{i} } } { { (E_{i} -
\mu_{i}) \exp \left\{ {{ E_{i} - \mu_{i} } \over T} \right\} }
\over { \left( \exp \left\{ {{ E_{i} - \mu_{i} } \over T} \right\}
+ g_{i} \right)^{2} } }\ , \label{entropy}
\end{equation}
\end{mathletters}

\noindent where $E_{i}= ( m_{i}^{2} + p^{2} )^{1/2}$ and $m_{i}$,
$B_{i}$, $S_{i}$, $\mu_{i}$, $s_{i}$ and $g_{i}$ are the mass,
baryon number, strangeness, chemical potential, spin and a
statistical factor of specie $i$ respectively (we treat an
antiparticle as a different specie). And $\mu_{i} = B_{i}\mu_{B} +
S_{i}\mu_{S}$, where $\mu_{B}$ and $\mu_{S}$ are overall baryon
number and strangeness chemical potentials respectively.

We shall work here within the usual timetable of the events in the
CRR of a given ion collision (for more details see e.g.
\cite{Blaizot:1989ec}). We fix $t=0$ at the moment of the maximal
overlap of the nuclei. After half of the time the nuclei need to
cross each other, matter appears in the CRR. We assume that soon
thereafter matter thermalizes and this moment, $t_{0}$, is
estimated at about 1 fm \cite{Blaizot:1989ec,Bjorken:1983qr}. Then
matter starts to expand and cool and after reaching the freeze-out
temperature it is no longer a thermodynamical system. We denote
this moment as $t_{f.o.}$. As we have already mentioned in the
introduction, this matter is the hadron gas, which consists of all
hadrons up to $\Omega^{-}$ baryon. The expansion proceeds
according to the relativistic hydrodynamic equations and for the
longitudinal component we have the following solution (for details
see e.g. \cite{Bjorken:1983qr,Cleymans:1986wb})

\begin{equation}
s(t)= { {s_{0}t_{0}} \over t } \;,\;\;\;\;\;\;\;\;\; n_{B}(t)= {
{n_{B}^{0}t_{0}} \over t } \ , \label{hydro}
\end{equation}

\noindent where $s_{0}$ and $n_{B}^{0}$ are initial densities of
the entropy and the baryon number respectively. The superimposed
transverse expansion has the form of the rarefaction wave moving
radially inward with a sound velocity $c_{s}$ and the transverse
flow which starts outside the wave
\cite{Bjorken:1983qr,Baym:1983sr}. But as it has been mentioned in
the introduction, since the temperature decreases rapidly outside
the wave (for the most important times for absorption processes),
we shall ignore possible $J/\Psi$ scattering there. It means that
$J/\Psi$ suppression is slightly underestimated here.

To obtain the time dependence of temperature and baryon number and
strangeness chemical potentials one has to solve numerically
equations (\ref{barnumb} - \ref{entropy})  with $s$, $n_{B}$ and
$n_{S}$ given as time dependent quantities. For $s(t)$, $n_{B}(t)$
we have expressions (\ref{hydro})  and $n_{S}=0$ since we put the
overall strangeness equal to zero during all the evolution (for
more details see \cite{Prorok:1995xz}).

The sound velocity squared is given by  $c_{s}^{2}= { {\partial P}
\over {\partial \epsilon} }$ and can be evaluated numerically
\cite{Prorok:1995xz,Prorok:2001ut}.

\section { J/$\Psi$ absorption in hadronic matter }

In a high energy heavy-ion collision, charmonium states are
produced mainly through gluon fusion and it takes place during the
overlap of colliding nuclei. For the purpose of our model, we
shall assume that all $c\bar{c}$ pairs are created at the moment
$t=0$. Before the fusion, gluons can suffer multiple elastic
scattering on nucleons and gain some additional transverse
momentum in this way
\cite{Hufner:1988wz,Gavin:1988tw,Blaizot:1989hh}. This manifests
for instance in the observed broadening of the $p_{T}$
distribution of $J/\Psi$ \cite{Badier:1983dg}. Following
\cite{Gupta:1992cd}, we express this effect by the transverse
momentum distribution of the charmonium states

\begin{equation} g(p_{T},\epsilon_{0}) = { {2p_{T}}
\over {\langle p_{T}^{2}\rangle_{J/\Psi}^{AB}(\epsilon_{0})} }
\cdot \exp \left\{- { { p_{T}^{2}} \over {\langle
p_{T}^{2}\rangle_{J/\Psi}^{AB}(\epsilon_{0})} } \right\}\ ,
\label{transv}
\end{equation}

\noindent where $\langle
p_{T}^{2}\rangle_{J/\Psi}^{AB}(\epsilon_{0})$ is the mean squared
transverse momentum of $J/\Psi$ gained in an A-B collision with
the initial energy density $\epsilon_{0}$. The momentum can be
expressed as (for details see \cite{Baglin:1991fy})

\begin{equation}
\langle p_{T}^{2}\rangle_{J/\Psi}^{AB}(\epsilon)=\langle
p_{T}^{2}\rangle_{J/\Psi}^{pp} + K \cdot \epsilon \ , \label{5}
\end{equation}

\noindent with $K=0.27$ fm$^{3}$GeV and $\langle
p_{T}^{2}\rangle_{J/\Psi}^{pp}=1.24$ GeV$^{2}$ taken from a fit to
the $J/\Psi$ data of NA38 Collaboration \cite{Baglin:1991fy}. The
expression in (\ref{transv})  is normalized to unity and is
treated as the initial momentum distribution of charmonium states
here.

For the simplicity of our model, we shall assume that all
charmonium states are completely formed and can be absorbed by the
constituents of a surrounding medium from the moment of creation.
It means that we neglect a whole complex process of $J/\Psi$
formation as presented in \cite{Kharzeev:1997yx,Satz:1997ib}. The
main feature of the above-mentioned process is that, soon after
the moment of production, the $c\bar{c}$ pair binds a soft gluon
and creates a pre-resonance $c\bar{c}-g$ state, from which, after
a time of the order of 0.3 fm, a physical charmonium state is
formed. This means that the possible nuclear absorption of
charmonium is, in fact, the absorption of the $c\bar{c}-g$ state.
But the latest has the cross-section $\sigma_{abs}=7.3$ mb, which
is much higher than $J/\Psi-Nucleon$ absorption cross-section
$\sigma_{\psi N} \cong 3-5$ mb obtained from p-A data
\cite{Gerschel:1988wn,Badier:1983dg,Gavin:1997yd}. This justifies
our assumption: taking into account $c\bar{c}-g$ absorption
instead of charmonium disintegration in the nuclear matter would
only strengthen $J/\Psi$ suppression.

According to the above assumption, charmonium states can be
absorbed first in the nuclear matter and soon later, when the
matter appears in the CRR, in the hadron gas. Since these two
processes are separate in time, $J/\Psi$ survival factor for a
heavy-ion collision with the initial energy density
$\epsilon_{0}$, may be written in the form

\begin{equation}
{\cal N}(\epsilon_{0}) = {\cal N}_{n.m.}(\epsilon_{0}) \cdot {\cal
N}_{h.g.}(\epsilon_{0})\ , \label{6}
\end{equation}

\noindent where ${\cal N}_{n.m.}(\epsilon_{0})$ and ${\cal
N}_{h.g.}(\epsilon_{0})$ are $J/\Psi$ survival factors in the
nuclear matter and the hadron gas, respectively. For ${\cal
N}_{n.m.}(\epsilon_{0})$ we have the usual approximation
\cite{Gavin:1997yd,Gerschel:1992uh,Gavin:1996en,Bella}

\begin{equation}
{\cal N}_{n.m.}(\epsilon_{0}) \cong \exp \left\{ -\sigma_{\psi N}
\rho_{0} L \right\}\ , \label{7}
\end{equation}

\noindent where $\rho_{0}$ is the nuclear matter density and $L$
the mean path length of the $J/\Psi$ through the colliding nuclei.
For the last quantity, we use the expression given in
\cite{Bella}:

\begin{equation}
 \rho_{0}L(b) =
{1 \over {2 T_{AB}} } \int d^{2}\vec{s}\; T_{A}(\vec{s})
T_{B}(\vec{s} - \vec{b}) \left[ {{A-1} \over A} T_{A}(\vec{s}) +
{{B-1} \over B} T_{B}(\vec{s} - \vec{b}) \right]\ , \label{8}
\end{equation}

\noindent where $T_{AB}(b) = \int d^{2}\vec{s}\; T_{A}(\vec{s})
T_{B}(\vec{s} - \vec{b})$, $T_{A}(\vec{s}) = \int dz
\rho_{A}(\vec{s},z)$ is the nuclear density profile function,
$\rho_{A}(\vec{s},z)$ the nuclear matter density distribution
(normalized to $A$) and $b$ the impact parameter. How to obtain
$\epsilon_{0}$ as a function of $b$ will be presented further.

To estimate ${\cal N}_{h.g.}(\epsilon_{0})$ we follow the idea
presented in \cite{Blaizot:1989ec}, but now generalized to the
case of the gas which consists of different species of particles.
We shall focus on the plane $z=0$ ($z$ is a collision axis) and
put $J/\Psi$ longitudinal momentum equal to zero. Now the
$p_{T}$-dependent $J/\Psi$ survival factor ${\cal
N}_{h.g.}(p_{T})$ is given by (for details see
\cite{Prorok:1994fb})

\begin{equation}
{\cal N}_{h.g.}(p_{T})= \int
d^{2}\vec{s} f_{0}(s,p_{T}) \exp \left\{ -\int_{t_{0}}^{t_{f}} dt
\sum_{i=1}^{l} \int { {d^{3}\vec{q}} \over {(2\pi)^{3}} }
f_{i}(\vec{q},t) \sigma_{i} v_{rel,i} { {p_{\nu}q_{i}^{\nu}} \over
{EE^{\prime}_{i}} } \right\}\ , \label{9}
\end{equation}

\noindent where the sum in the power is over all taken species of
scatters (hadrons), $p^{\nu}=(E,\vec{p}_{T})$ and
$q_{i}^{\nu}=(E^{\prime}_{i},\vec{q})$ are four momenta of
$J/\Psi$ and hadron specie $i$ respectively, $\vec{v}=
\vec{p}_{T}/E$ is the velocity of the former, $\sigma_{i}$ states
for the absorption cross-section of $J/\Psi-h_{i}$ scattering and
$v_{rel,i}$ is the relative velocity of $h_{i}$ hadron with
respect to $J/\Psi$. When $M$ and $m_{i}$ denote $J/\Psi$ and
$h_{i}$ masses, respectively ($M= 3097$ MeV), $v_{rel,i}$ reads

\begin{equation}
v_{rel,i}=\left( 1- { {m_{i}^{2}M^{2}} \over
{(p_{\nu}q_{i}^{\nu})^{2}} } \right)^{{1 \over 2}} \ . \label{10}
\end{equation}

The upper limit of the time integral in (\ref{9}) , $t_{f}$, is
equal to $t_{f.o.}$ or to $t_{esc}$ -- the moment of leaving by a
given $J/\Psi$ of the hadron medium, if the final-size effects are
considered and $t_{esc} < t_{f.o.}$. For $\sigma_{i}$ we have
assumed that it equals zero for $(p^{\nu}+q_{i}^{\nu})^{2} <
(2m_{D} + m_{X})^{2}$ and is constant elsewhere ($m_{D}$ is a
charm meson mass, $m_{D}= 1867$ MeV). For hadron specie $i$ we
have usual Bose-Einstein or Fermi-Dirac distribution (we neglect
any possible spatial dependence here)

\begin{equation}
f_{i}(\vec{q},t)=f_{i}(q,t)={ {2s_{i}+1} \over {
\exp \left\{ { { E^{\prime}_{i}-\mu_{i}(t)} \over {T(t)} }
\right\} + g_{i} } }\ . \label{11}
\end{equation}

In the following, we shall consider only $J/\Psi$ initial
distribution $f_{0}(s,p_{T})$ that factorizes into
$f_{0}(s)g(p_{T})$ and the momentum distribution $g(p_{T})$ will
be given by (\ref{transv}) . We assume at the first step that the
transverse size of the hadron medium is much greater than
$t_{f.o.}$ and also much greater than the size of the area where
$f_{0}(s)$ is non-zero. Additionally we assume that $f_{0}(s)$ is
uniform and normalized to unity. Note that the first assumption
overestimates the suppression but the second, in the presence of
the first, has no any calculable effect here. As a result, ${\cal
N}_{h.g.}(p_{T})$ simplifies to

\begin{equation}
{\cal N}_{h.g.}(p_{T}) = g(p_{T},\epsilon_{0}) \cdot \exp \left\{
-\int_{t_{0}}^{t_{f.o.}} dt \sum_{i=1}^{l} \int { {d^{3}\vec{q}}
\over {(2\pi)^{3}} } f_{i}(\vec{q},t) \sigma_{i} v_{rel,i} {
{p_{\nu}q_{i}^{\nu}} \over {EE^{\prime}_{i}} } \right\}\ ,
\label{12}
\end{equation}

\noindent To obtain ${\cal N}_{h.g.}(\epsilon_{0})$ one needs only
to integrate (\ref{12})  over $p_{T}$:

\begin{equation}
{\cal N}_{h.g.}(\epsilon_{0}) = \int dp_{T}\;
g(p_{T},\epsilon_{0}) \cdot \exp \left\{ -\int_{t_{0}}^{t_{f.o.}}
dt \sum_{i=1}^{l} \int { {d^{3}\vec{q}} \over {(2\pi)^{3}} }
f_{i}(\vec{q},t) \sigma_{i} v_{rel,i} { {p_{\nu}q_{i}^{\nu}} \over
{EE^{\prime}_{i}} } \right\}\ . \label{13}
\end{equation}

Now we would like to take the final-size effects and the
transverse expansion into account in our model. To do this
directly, we would have to come back to the formula given by
(\ref{9}) and integrate it, instead of (\ref{12}) , over $p_{T}$.
But this would involve a five-dimensional integral (the
three-dimensional integral over $\vec{q}$ simplifies to the
one-dimensional one, in fact) instead of the three-dimensional
integral of (\ref{13}). Therefore, we need to simplify in some way
the direct method just mentioned above. We shall define an average
time of leaving the hadron medium by $J/\Psi$'s with the velocity
$v$ produced in an A-B collision at impact parameter $b$, $\langle
t_{esc}\rangle(b,v)$. Then, if this quantity is less than
$t_{f.o.}$, we will put it instead of $t_{f.o.}$ as the upper
limit of the integral over $t$ in (\ref{13}).

\begin{figure}
\begin{center}{
{\epsfig{file=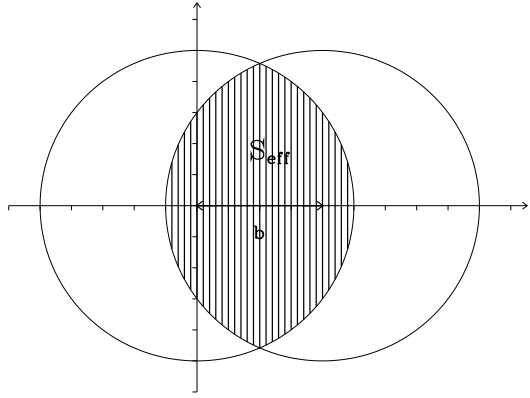,width=7cm}} }\end{center} \caption{View
of a Pb-Pb collision at impact parameter $b$ in the transverse
plane ($z=0$). The region where the nuclei overlap has been
hatched and its area equals $S_{eff}$.}
\label{Fig.2.}
\end{figure}

Let us consider an A-B collision at impact parameter $b$. Since we
will compare final results with the latest data of NA50 which are
for Pb-Pb collisions \cite{Abr50}, we focus on the case of A=B
here. So, for the collision at impact parameter $b$ we have the
situation in the plane $z=0$ as presented in Fig.\,\ref{Fig.2.},
where $S_{eff}$ means the area of the overlap of the colliding
nuclei. We shall assume here, that the hadron medium, which
appears in the space between the nuclei after they crossed each
other also has the shape of $S_{eff}$ at $t_{0}$ in the plane
$z=0$. And additionally, the transverse expansion  starts in the
form of the rarefaction wave moving inward $S_{eff}$ at $t_{0}$.
Then, for a $J/\Psi$ which is at $\vec{r} \in S_{eff}$ at the
moment $t_{0}$ and has the velocity $\vec{v}$ we denote by
$t_{esc}$ the moment of crossing the border of the hadron gas. It
means that $t_{esc}$ is a solution of the equation $\mid \vec{d} +
\vec{v} (t-t_{0}) \mid = R_{A} - c_{s} (t-t_{0})$, where $R_{A} =
r_{0} \cdot A^{{1 \over 3}}$ is the nucleus radius and $\vec{d} =
\vec{r} - \vec{b}$ for the angel between $\vec{r}$ and $\vec{v}$
such that the $J/\Psi$ will cross this part of the edge of the
area of the hadron gas which was created by the projectile and
$\vec{d} = \vec{r}$ in the opposite. Having obtained $t_{esc}$, we
average it over the angel between $\vec{r}$ and $\vec{v}$, i.e. we
integrate $t_{esc}$ over this angel and divide by $2\pi$. Then we
average the result over $S_{eff}$ with the weight given by

\begin{equation}
p_{J/\Psi}(\vec{r}) = {
{T_{A}(\vec{r})T_{B}(\vec{r} - \vec{b})} \over {T_{AB}(b)} }
\label{14}
\end{equation}

\noindent and we obtain $\langle t_{esc}\rangle(b,v)$. So, the
final expression for ${\cal N}_{h.g.}(\epsilon_{0})$ when the
transverse expansion is taken into account reads

\begin{equation}
{\cal N}_{h.g.}(\epsilon_{0}) = \int dp_{T}\;
g(p_{T},\epsilon_{0}) \cdot \exp \left\{ -\int_{t_{0}}^{t_{final}}
dt \sum_{i=1}^{l} \int { {d^{3}\vec{q}} \over {(2\pi)^{3}} }
f_{i}(\vec{q},t) \sigma_{i} v_{rel,i} { {p_{\nu}q_{i}^{\nu}} \over
{EE^{\prime}_{i}} } \right\}\ , \label{15}
\end{equation}

\noindent where $t_{final}=min\{ \langle
t_{esc}\rangle,t_{f.o.}\}$.

\section { The energy density in the CRR }

We compare our theoretical estimations for $J/\Psi$ survival
factor with the experimental data \cite{Abr50} presented as a
function of $\epsilon_{0}$. Usually, this quantity is estimated
from the well-known Bjorken formula

\begin{equation}
\epsilon_{0} = { {3 \cdot E_{T}} \over {
\Delta\eta S_{eff} t_{0} } }\ , \label{16}
\end{equation}

\noindent where $\Delta\eta$ is the pseudo-rapidity range and
$E_{T}$ is the neutral transverse energy.

In further considerations we will need the formula for the number
of participating nucleons as a function of impact parameter $b$,
which is given by the rough approximation (commonly used in the
early nineties)

\begin{equation}
N_{part}(b) = \int_{S_{eff}} d^{2}\vec{s} \left\{ T_{A}(\vec{s}) +
T_{B}(\vec{s} - \vec{b}) \right\}\ , \label{17}
\end{equation}

\noindent or in the term of the number of "wounded" nucleons
\cite{Kharzeev:1997yx,Bella}

\begin{equation}
N_{wound}(b) = \int d^{2}\vec{s}\; T_{A}(\vec{s}) \left\{ 1 -
\left[ 1-{{\sigma_{N}} \over B}T_{B}(\vec{s} - \vec{b})
\right]^{B}\right\} + \int d^{2}\vec{s}\; T_{B}(\vec{s} - \vec{b})
\left\{ 1 - \left[ 1-{{\sigma_{N}} \over A}T_{A}(\vec{s})
\right]^{A}\right\}\ . \label{nwound}
\end{equation}

\begin{figure}[htb]
\begin{minipage}[t]{75mm}
{\psfig{figure= 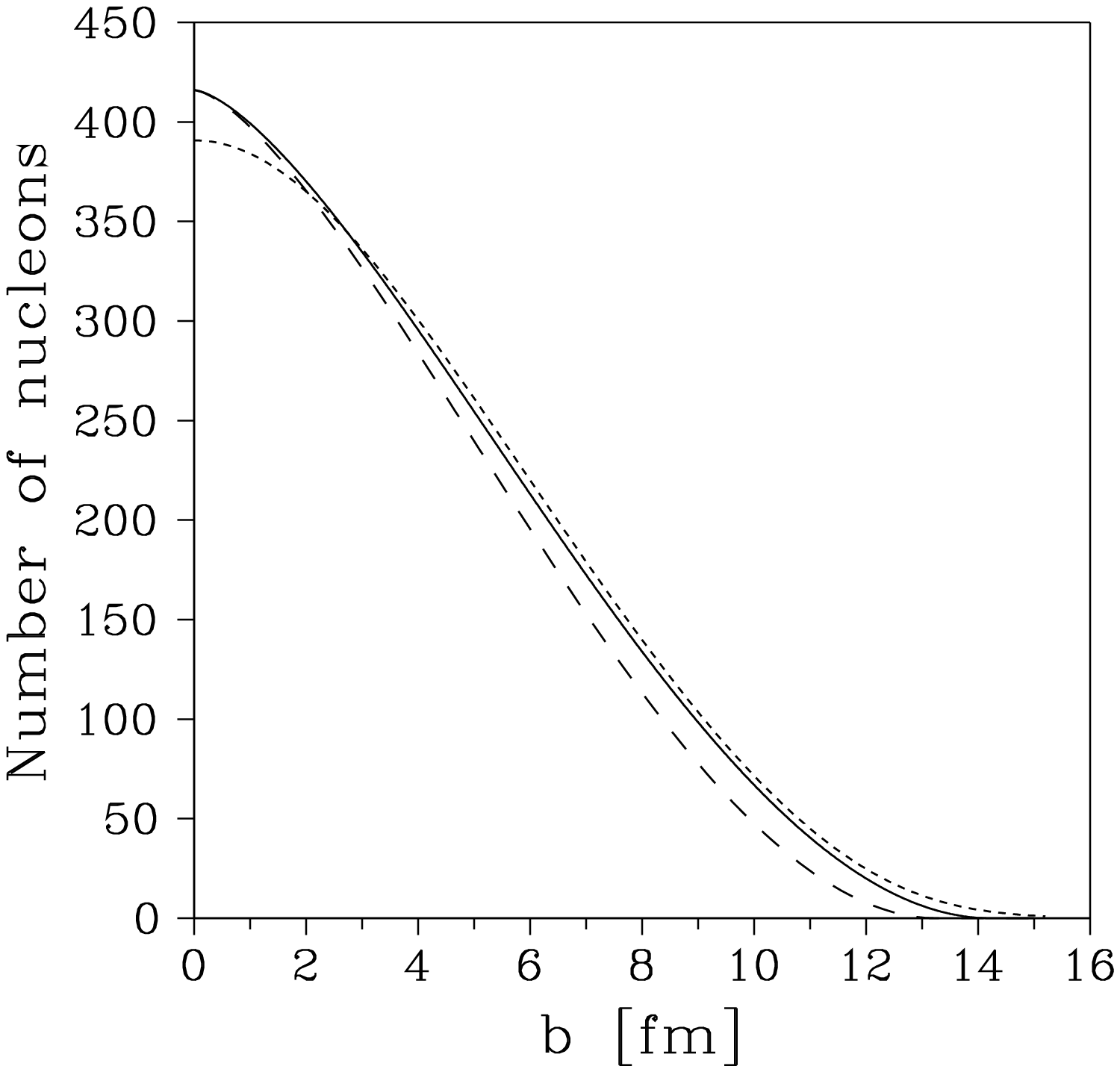,height=\textwidth,width=\textwidth}}
\caption{Number of participating nucleons as a function of $b$ for
a Pb-Pb collision, estimated as:$\; N_{part}(b)$ for the uniform
nuclear matter density and $r_{0}=1.2$ fm (solid), $r_{0}=1.12$ fm
(dashed); $\; N_{wound}(b)$ for the Woods-Saxon distribution
(short-dashed).} \label{Fig.3.}
\end{minipage}
\hspace{\fill}
\begin{minipage}[t]{75mm}
{\psfig{figure= 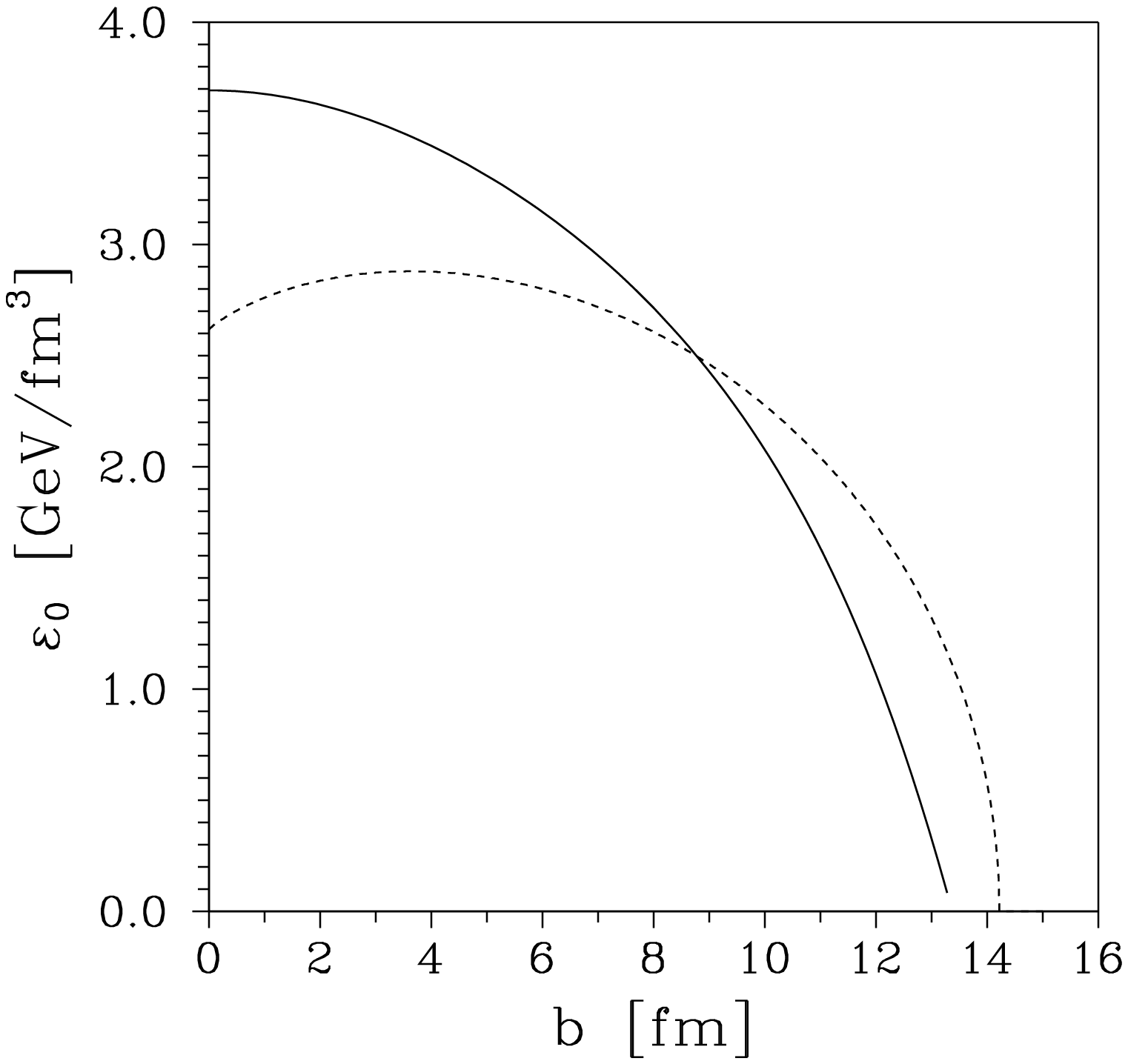 ,height=\textwidth,width=\textwidth}}
\caption{The initial energy density $\epsilon_0$ in the CRR for
Pb-Pb collisions as the function of impact parameter $b$ extracted
from the NA50 data \protect\cite{Abr50} (solid) and obtained from
(\ref{etpart}) (short-dashed).} \label{Fig.4.}
\end{minipage}
\end{figure}

\noindent The both expressions are depicted in Fig.\,\ref{Fig.3.}.
Note that $N_{part}(b)$ is estimated for the uniform nuclear
matter density with $r_{0}=1.2$ fm and $r_{0}=1.12$ fm, whereas
$N_{wound}(b)$ is evaluated for the Woods-Saxon nuclear matter
density distribution with parameters taken from \cite{Jager}. With
the use of $N_{wound}(b)$, the relation between $E_{T}$ and $b$ is
established \cite{Kharzeev:1997yx,Bella}

\begin{equation}
E_{T}= q \cdot N_{wound}(b)\ , \label{etwound}
\end{equation}

\noindent where $q=0.4$ GeV. We could put (\ref{etwound}) into
(\ref{16}) to obtain the dependence between $\epsilon_{0}$ and $b$
but the ratio ${ {N_{wound}(b)} \over {S_{eff}(b)}}$ is divergent
when $b \rightarrow R_{A}+R_{B}$ ($R_{A}$ and $R_{B}$ are radii of
a projectile and a target, respectively). From Fig.\,\ref{Fig.3.}
we can see that in the case of Pb-Pb collisions, $N_{wound}(b)$ do
not differ substantially from $N_{part}(b)$ with $r_{0}=1.2$ fm
(besides the low $b$ region). Therefore we can assume that for
Pb-Pb collisions of NA50 the following approximation is valid:

\begin{equation}
E_{T} \cong q \cdot N_{part}\ . \label{etappr}
\end{equation}

Having put (\ref{etappr})  into (\ref{16})  we obtain
$\epsilon_{0}$ as a function of $b$

\begin{equation}
\epsilon_{0}(b) = { {N_{part}(b)} \over {S_{eff}(b)} }\ ,
\label{etpart}
\end{equation}
\noindent where we have also used the value $\Delta\eta=1.2$ of
NA50 \cite{Abr50}. The above function is depicted in
Fig.\,\ref{Fig.4.}, together with $\epsilon_{0}(b)$ obtained from
$\epsilon_{0}(E_{T})$ with the use of (\ref{etwound}). The
dependence of $\epsilon_{0}$ on $E_{T}$ has been extracted
directly from NA50 data \cite{Abr50}.

We shall later see that the low $b$ region (central collisions) is
crucial for the understanding of experimental data. Also different
plots in Fig.\,\ref{Fig.4.} will lead to different ranges of
initial energy densities obtained from formula (\ref{etpart}) or
given by NA50 Collaboration \cite{Abr50}.

\section { Results }

To evaluate formulae (\ref{13})  and (\ref{15})  we have to know
$T(t)$, $\mu_{B}(t)$ and $\mu_{S}(t)$ and how to obtain these
functions was explained in Sect.~\ref{hadgas}. But to follow all
that procedure we need initial values $s_{0}$ and $n_{B}^{0}$. To
estimate initial baryon number density $n_{B}^{0}$ we can use
experimental results for S-S \cite{Baechler:1991pd} or Au-Au
\cite{Stachel:1999rc,Ahle:1998jc} collisions. In the first
approximation we can assume that the baryon multiplicity per unit
rapidity in the CRR is proportional to the number of participating
nucleons. For a sulphur-sulphur collision we have $dN_{B}/dy \cong
6$ \cite{Baechler:1991pd} and 64 participating nucleons. For the
central collision of lead nuclei we can estimate the number of
participating nucleons at $2A = 416$, so we have $dN_{B}/dy \cong
39$. Having taken the initial volume in the CRR equal to $\pi
R_{A}^{2} \cdot 1$ fm, we arrive at $n_{B}^{0} \cong 0.25$
fm$^{-3}$. This is some underestimation because the S-S collisions
were at a beam energy of 200 GeV/nucleon, but Pb-Pb at 158
GeV/nucleon. From the Au-Au data extrapolation one can estimate
$n_{B}^{0} \cong 0.65$ fm$^{-3}$ \cite{Stachel:1999rc}. These
values are for central collisions, and for the higher impact
parameter (a more peripheral collision) the initial baryon number
density should be much lower. So, to simplify numerical
calculations we will keep $n_{B}^{0}$ constant over the all range
of $b$ and additionally, to check the possible dependence on
$n_{B}^{0}$, we will do our estimations for $n_{B}^{0}$
substantially lower, i.e. $n_{B}^{0} = 0.05$ fm$^{-3}$.

Now, to find $s_{0}$, first we have to solve (\ref{energy} -
\ref{strange}) with respect to $T$, $\mu_{S}$ and $\mu_{B}$, where
we put $\epsilon = \epsilon_{0}$, $n_{B} = n_{B}^{0}$ and
$n_{S}=0$. Then, having put $T$, $\mu_{S}$ and $\mu_{B}$ into
(\ref{entropy})  we obtain $s_{0}$. Finally, expressing left sides
of (\ref{barnumb},\ref{entropy}) by (\ref{hydro})  and after then
solving (\ref{barnumb} - \ref{entropy})  numerically we can obtain
$T$, $\mu_{S}$ and $\mu_{B}$ as functions of time. In fact,
evaluating formulae (\ref{13})  and (\ref{15})  we do the
following: first, we calculate $T=T(t)$ which turns out to be very
well approximated by the expression

\begin{equation}
T(t) \cong T_{0} \cdot t^{-a} \label{temperature}
\end{equation}

\noindent and then we put this approximation into (\ref{13})  and
(\ref{15}) . And for $\mu_{S}(t)$ and $\mu_{B}(t)$ in
$f_{i}(\vec{q},t)$ we put solutions of
(\ref{barnumb},\ref{strange}) where $n_{B}$ is given by
(\ref{hydro}), $T$ by (\ref{temperature}) and $n_{S}=0$. But the
exponent $a$ in (\ref{temperature})  has proven not to be unique
for the whole range of $T_{0}$ considered here. One gets different
values of the initial energy density $\epsilon_0$ for different
values of the impact parameter $b$  and for different geometry of
the collision process. So $b$ dependent $a$ gives also $b$
dependent freeze-out time $t_{f.o}$. The density $\epsilon_{0}$ is
extracted from the dependence represented by the solid line in
Fig.\,\ref{Fig.4.} for different values of $b$ and Eqs.
(\ref{energy} - \ref{strange}) are solved. We have evaluated the
suppression factor up to $\epsilon_{0} = 3.7$ GeV/fm$^{3}$. This
gives the maximal possible $T_{0}$, $T_{0,max}$, equal to 221.8
MeV (for $n_{B}^{0} = 0.65$ fm$^{-3}$), 226 MeV (for $n_{B}^{0} =
0.25$ fm$^{-3}$) or 226.7 MeV (for $n_{B}^{0} = 0.05$ fm$^{-3}$).

This procedure allows to evaluate $J/\Psi$ survival factor given
by (\ref{13}) . Because of the lack of data, we shall assume only
two types of the cross-section, the first, $\sigma_{b}$, for
$J/\Psi$-baryon scattering and the second, $\sigma_{m}$, for
$J/\Psi$-meson scattering. For $\sigma_{b}$ we put $\sigma_{b} =
\sigma_{J/\psi N}$. As far as $\sigma_m$ is concerned, we assume
that this cross-section is $2/3$ of the corresponding cross
section for baryons, which is due to the quark counting. In the
following, we will use values of $J/\Psi-Nucleon$ absorption
cross-section $\sigma_{J/\psi N} \cong 3-5$ mb obtained from p-A
data \cite{Gerschel:1988wn,Badier:1983dg,Gavin:1997yd}. At the
beginning, to illustrate how the value of power $a$ influences
$J/\Psi$ suppression we present in Fig.\,\ref{Fig.1.} two results:
the first for $a={1 \over 3}$ (what is the exact value for a free
massless gas) and the second for $a={1 \over {5.6}}$ (what is the
approximate value for the hadron gas and $T_{0} \cong 200$ MeV).
We can see that the suppression improves more than twice for the
highest $\epsilon_{0}$ indeed.

To make our investigations more realistic we have to take into
account that only about $60 \%$ of $J/\Psi$ measured are directly
produced during collision. The rest is the result of $\chi$ ($\sim
30 \%$) and $\psi'$ ($\sim 10 \%$) decay \cite{Satz:1997ib}.
Therefore the realistic $J/\Psi$ survival factor should read

\begin{equation}
{\cal N}(\epsilon_{0})=0.6{\cal
N}_{J/\psi}(\epsilon_{0})+0.3{\cal N}_{\chi}(\epsilon_{0})+
0.1{\cal N}_{\psi'}(\epsilon_{0})\;, \label{survsum}
\end{equation}

\noindent where ${\cal N}_{J/\psi}(\epsilon_{0})$, ${\cal
N}_{\chi}(\epsilon_{0})$ and ${\cal N}_{\psi'}(\epsilon_{0})$ are
given also by formulae (\ref{transv}-\ref{15})  but with $\langle
p_{T}^{2}\rangle_{J/\Psi}^{AB}(\epsilon)=\langle
p_{T}^{2}\rangle_{J/\Psi}^{AB}(\epsilon), \langle
p_{T}^{2}\rangle_{\chi}^{AB}(\epsilon), \langle
p_{T}^{2}\rangle_{\psi'}^{AB}(\epsilon)$, $K_{J/\psi}=K_{J/\psi},
K_{\chi}, K_{\psi'}$, $\sigma_{J/\psi N}=\sigma_{J/\psi N},
\sigma_{\chi N}, \sigma_{\psi' N}$ and $M=M_{J/\psi}, M_{\chi},
M_{\psi'}$ respectively. The remaining problem is whether formula
(\ref{5})  is valid for $\chi$ and $\psi'$. There are data for
$\langle p_{T}^{2}\rangle_{\psi'}^{PbPb}$ \cite{Abreu:1998vw} and
they shows that $\langle p_{T}^{2}\rangle_{\psi'}^{PbPb} \approx
1.4 \langle p_{T}^{2}\rangle_{J/\Psi}^{PbPb}$. So, we assume that
the above is also true for $\langle
p_{T}^{2}\rangle_{\psi'}^{AB}(\epsilon)$, i.e.

\begin{equation}
\langle p_{T}^{2}\rangle_{\psi'}^{AB}(\epsilon)=1.4 \langle
p_{T}^{2}\rangle_{J/\Psi}^{AB}(\epsilon) \label{pt}
\end{equation}

\noindent with $\langle p_{T}^{2}\rangle_{J/\Psi}^{AB}(\epsilon)$
given by (\ref{5}) . For $\chi$ we believe that the inequality

\begin{equation}
\langle p_{T}^{2}\rangle_{J/\Psi}^{AB}\;\; \leq \;\; \langle
p_{T}^{2}\rangle_{\chi}^{AB}\;\; \leq \;\; \langle
p_{T}^{2}\rangle_{\psi'}^{AB} \label{ineq}
\end{equation}

\noindent should be valid and therefore assume that (\ref{pt})  is
true also in this case. Anyway, the exact form of $\langle
p_{T}^{2}\rangle_{\chi}^{AB}(\epsilon)$ or $\langle
p_{T}^{2}\rangle_{\psi'}^{AB}(\epsilon)$ is not very important
because we checked that the suppression depends on this form very
weakly. First, we put $K_{J/\psi} = 0$ and the resulting $J/\Psi$
survival factor (for direct $J/\Psi$'s) differs only a few percent
for the highest $\epsilon_{0}$ from the one calculated with
formula (\ref{5})  unchanged. Second, when we use expression
(\ref{5}) also for $\chi$ and $\psi'$, the evaluated suppression
factor is the same as that calculated with the use of (\ref{pt}),
as far as plots are concerned.

\begin{figure}[htb]
\begin{minipage}[t]{75mm}
{\psfig{figure= 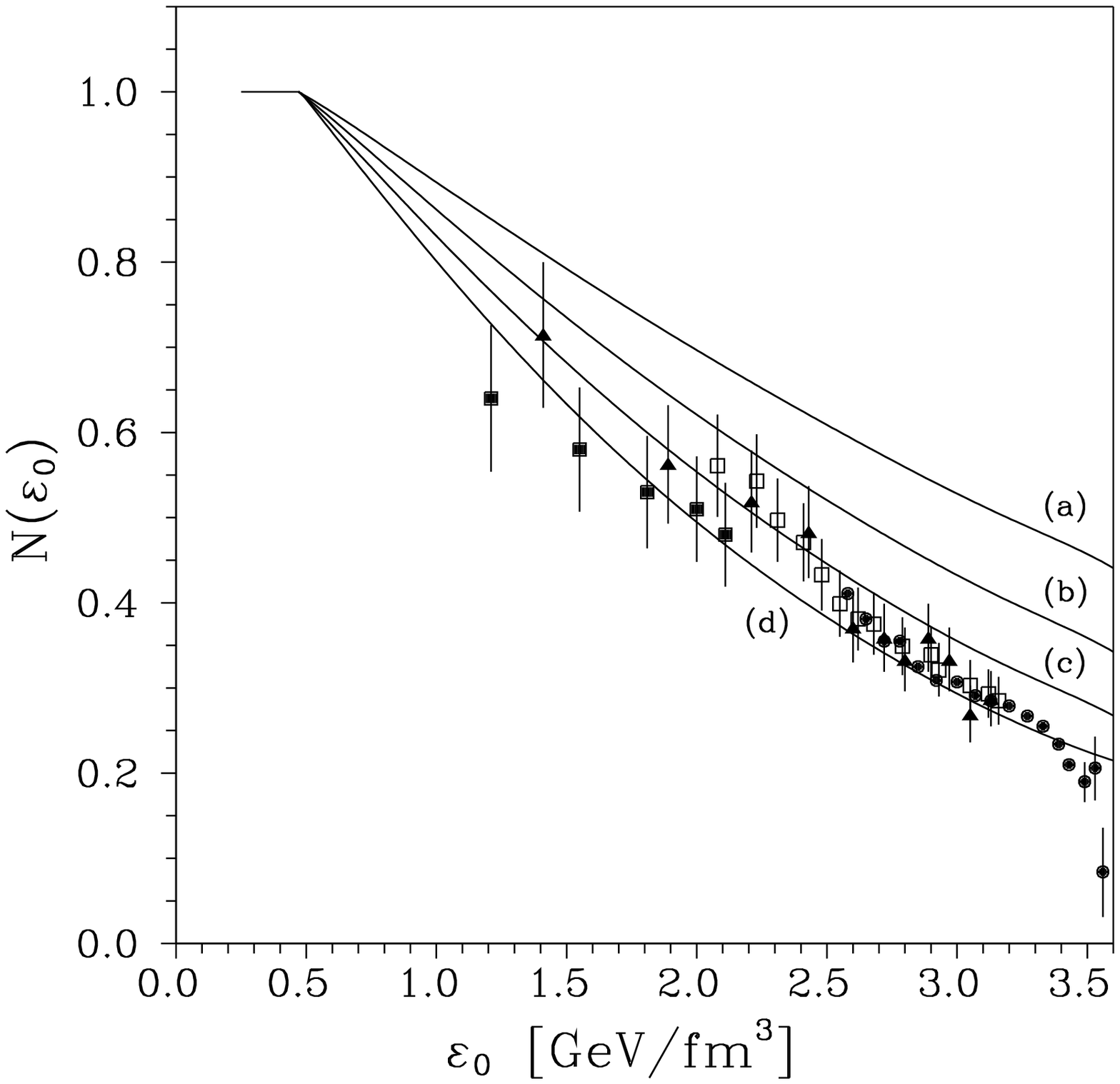,height=\textwidth,width=\textwidth}}
\caption{$J/\Psi$ suppression in the longitudinally expanding
hadron gas with the "infinite" transverse size and for
$n_{B}^{0}=0.25$ fm$^{-3}$ and $T_{f.o.}=140$ MeV: (a)
$\sigma_{b}=3$ mb, $\sigma_{m}=2$ mb; (b) $\sigma_{b}= 4$ mb,
$\sigma_{m}=2.66$ mb; (c) $\sigma_{b}=5$ mb, $\sigma_{m}=3.33$ mb;
(d) $\sigma_{b}=6$ mb, $\sigma_{m}=4$ mb . The black squares
correspond to the NA38 S-U data \protect\cite{Abr}, the black
triangles correspond to the 1996 NA50 Pb-Pb data, the white
squares to the 1996 analysis with minimum bias and the black
points to the 1998 analysis with minimum bias
\protect\cite{Abr50}. } \label{Fig.5.}
\end{minipage}
\hspace{\fill}
\begin{minipage}[t]{75mm}
{\psfig{figure= 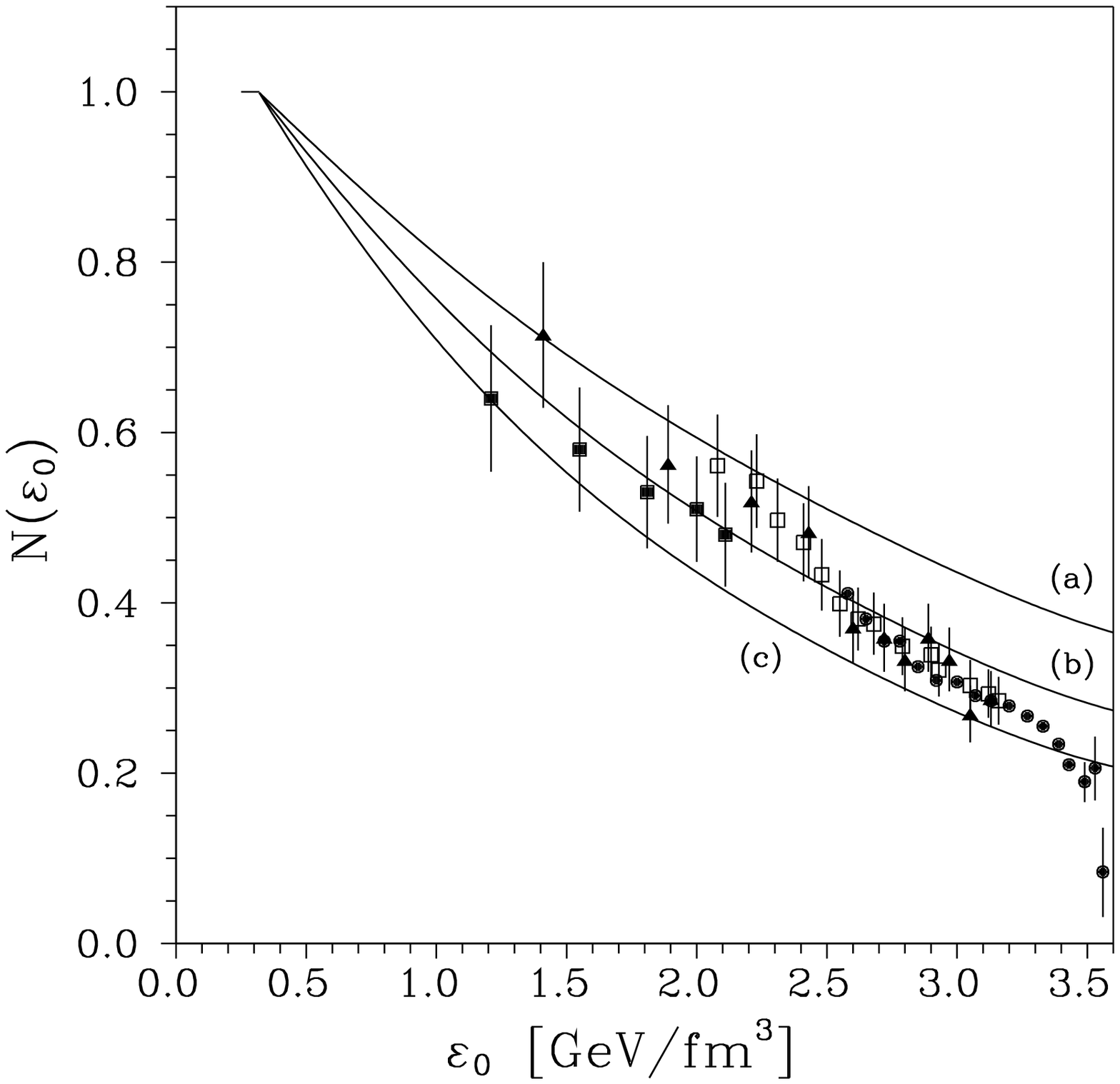 ,height=\textwidth,width=\textwidth}}
\caption{Same as Fig.\,\ref{Fig.5.} (except case (d), which is not
presented here) but for $T_{f.o.}=100$ MeV.} \label{Fig.6.}
\end{minipage}
\end{figure}

\begin{figure}[htb]
\begin{minipage}[t]{75mm}
{\psfig{figure= 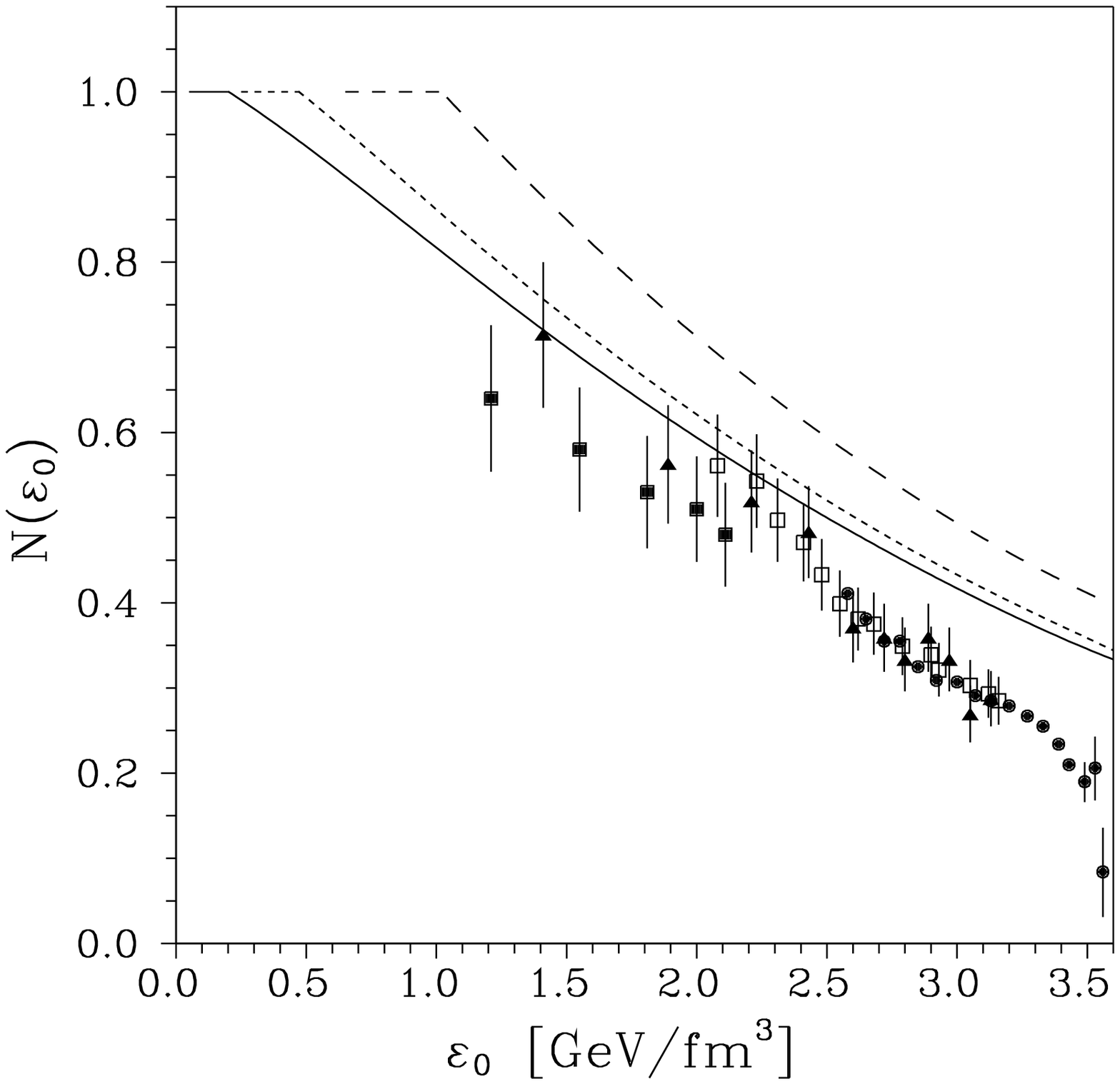,height=\textwidth,width=\textwidth}}
\caption{$J/\Psi$ suppression in the longitudinally expanding
hadron gas with the "infinite" transverse size and for
$\sigma_{b}=4$ mb, $\sigma_{m}=2.66$ mb and $T_{f.o.}=140$ MeV.
The curves correspond to $n_{B}^{0}=0.05$ fm$^{-3}$ (solid),
$n_{B}^{0}=0.25$ fm$^{-3}$ (short-dashed) and $n_{B}^{0}=0.65$
fm$^{-3}$ (dashed).The black squares represent the NA38 S-U data
\protect\cite{Abr}, the black triangles represent the 1996 NA50
Pb-Pb data, the white squares the 1996 analysis with minimum bias
and the black points the 1998 analysis with minimum bias
\protect\cite{Abr50}.} \label{Fig.7.}
\end{minipage}
\hspace{\fill}
\begin{minipage}[t]{75mm}
{\psfig{figure= 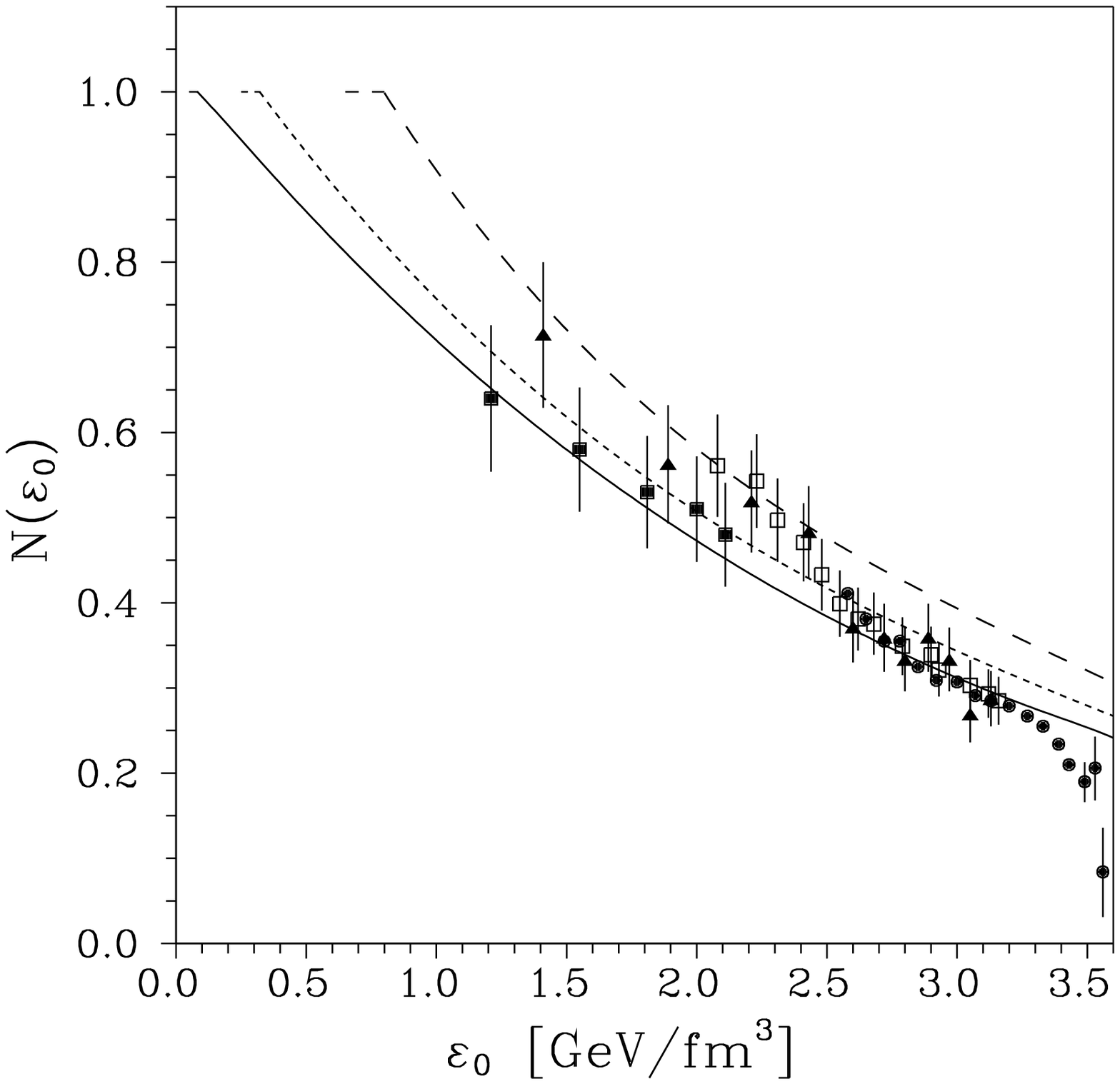 ,height=\textwidth,width=\textwidth}}
\caption{Same as Fig.\,\ref{Fig.7.} but for $T_{f.o.}=100$ MeV.}
\label{Fig.8.}
\end{minipage}
\end{figure}

To complete our estimations we need also values of cross-sections
for $\chi-baryon$ and $\psi'-baryon$ scatterings (we will still
hold that $\chi(\psi')-meson$ cross-section is ${2 \over 3}$ of
$\chi(\psi')-baryon$ cross-section). Since $J/\Psi$ is smaller
than $\chi$ or $\psi'$, $\chi-baryon$ and $\psi'-baryon$
cross-sections should be greater than $J/\Psi-baryon$ one. For
simplicity, we assume that all these cross-sections are equal.
This means that we {\it underestimate} $J/\Psi$ suppression, here.
The final results of calculations of (\ref{13}) are presented in
Figs.\,\ref{Fig.5.}-\ref{Fig.8.} for various sets of parameters of
our model (which are $T_{f.o.}, n_{B}^{0}, \sigma_{b}$). We
performed these calculations for two values of $T_{f.o.}=100, 140$
MeV which agree fairly well with values deduced from hadron yields
\cite{Stachel:1999rc}. For comparison, also the experimental data
are shown in Figs.\,\ref{Fig.5.}-\ref{Fig.8.}. The experimental
survival factor is defined as

\begin{equation}
{\cal N}_{exp}={ { {B_{\mu\mu}\sigma_{J/\psi}^{AB}}\over
{\sigma_{DY}^{AB}} } \over { {B_{\mu\mu}\sigma_{J/\psi}^{pp}}
\over {\sigma_{DY}^{pp}}  } }\;\;, \label{survexp1}
\end{equation}

\noindent where ${B_{\mu\mu}\sigma_{J/\psi}^{AB(pp)}} \over
{\sigma_{DY}^{AB(pp)}}$ is the ratio of the $J/\Psi$ to the
Drell-Yan production cross-section in A-B(p-p) interactions times
the branching ratio of the $J/\Psi$ into a muon pair. The values
of the ratio for p-p, S-U and Pb-Pb are taken from
\cite{Abr,Abr50}. Note that since the equality
$\sigma_{DY}^{AB}=\sigma_{DY}^{pp} \cdot AB$ has been confirmed
experimentally up to now \cite{Abreu:1998vw}, formula
(\ref{survexp1}) reduces to

\begin{equation}
{\cal N}_{exp}= { {\sigma_{J/\psi}^{AB}} \over {AB
\sigma_{DY}^{pp}} }\;\;, \label{survexp2}
\end{equation}

\noindent which is also given as the experimental survival factor,
for instance, in \cite{Lourenco:1996wn}.

Coming back to examination of our first results presented in
Figs.\,\ref{Fig.5.}-\ref{Fig.8.}, we can see that we have obtained
patterns of suppression which agree with the experimental data
fairly well for some values of parameters of our model. The data
prefer $\sigma_{b}=5-6$ mb and (or) $T_{f.o.}$ closer to 100 MeV.
Note that the dependence on the initial baryon number density is
substantial but for higher values of $n_{B}^{0}$, rather. The
lower the initial baryon number density, the deeper the
suppression. There are two reasons for such a behaviour: the
first, for the higher baryon number density, there are less
non-strange heavier mesons $\rho$, $\omega$ in the hadron gas of
the same $\epsilon_{0}$, but these particles create the most
weighty fraction of scatters, for which reaction (\ref{psiabs})
have no threshold at all; the second, the freeze-out time
$t_{f.o.}$ decreases with increasing $n_{B}^{0}$ for a given
$\epsilon_{0}$ in our model. For instance, for $\epsilon_{0}=3.5$
GeV/fm$^{3}$ and $T_{f.o.}=140$ MeV we have
$a=0.172,\;0.175,\;0.183$ and $t_{f.o.} = 15.7,\;14.7,\;11.6$ fm
for $n_{B}^{0}= 0.05,\;0.25,\;0.65$ fm$^{-3}$, respectively. We
can see also that the value $\sigma_{b}=3$ mb is too small to
obtain results comparable with the data, so we will leave aside
this value in further investigations.

\begin{figure}[htb]
\begin{minipage}[t]{75mm}
{\psfig{figure= 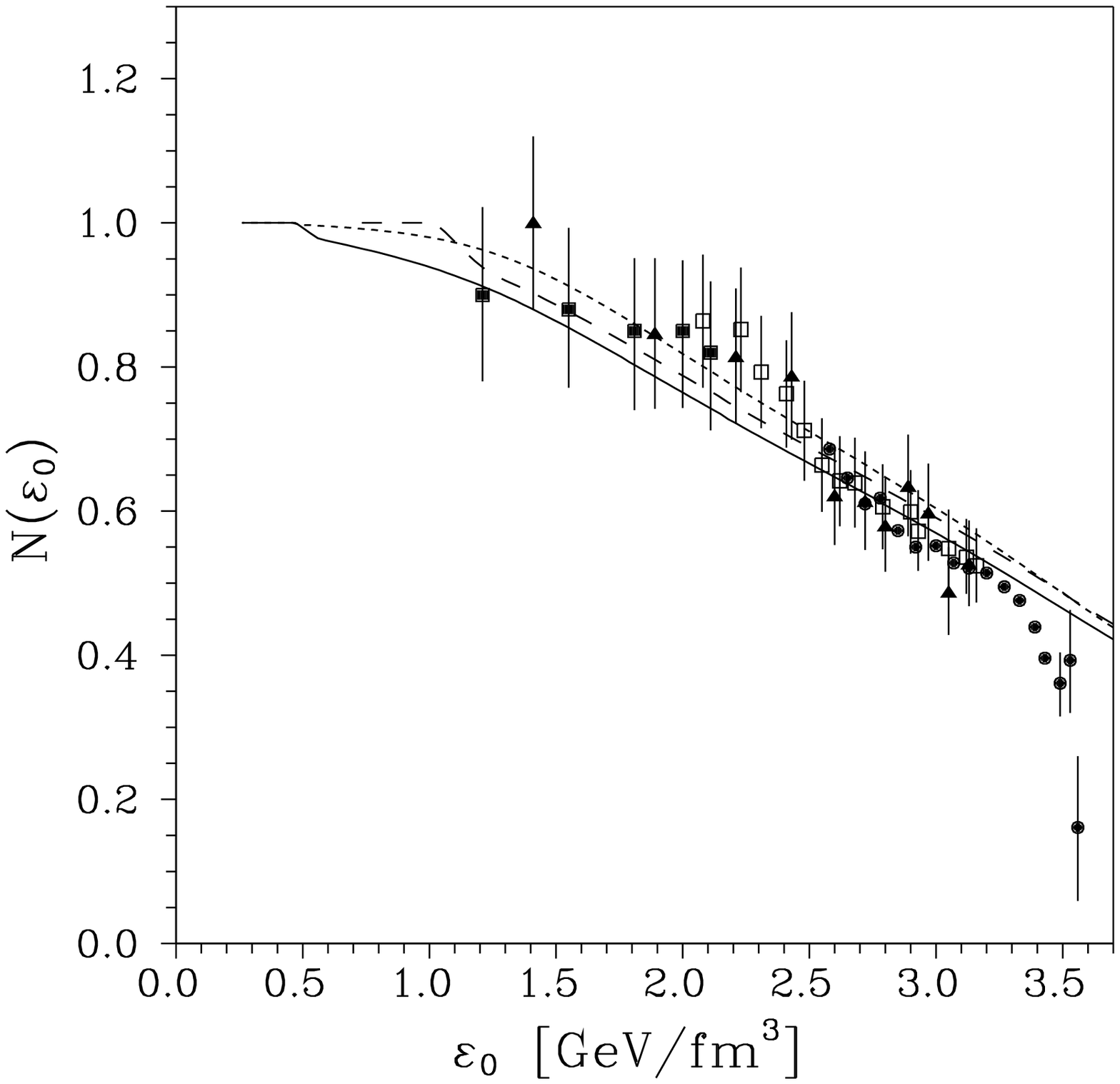,height=\textwidth,width=\textwidth}}
 \caption{$J/\Psi$ suppression in the longitudinally and
transversely expanding hadron gas for the Woods-Saxon nuclear
matter density distribution and $\sigma_{b}=4$ mb,
$\sigma_{m}=2.66$ mb and $T_{f.o.}=140$ MeV. The curves correspond
to $n_{B}^{0}=0.25$ fm$^{-3}$, $c_{s}=0.45$, $r_{0}=1.2$ fm
(solid), $n_{B}^{0}=0.65$ fm$^{-3}$, $c_{s}=0.46$, $r_{0}=1.2$ fm
(dashed) and  $n_{B}^{0}=0.25$ fm$^{-3}$, $c_{s}=0.45$,
$r_{0}=1.12$ fm (short-dashed). The black squares represent the
NA38 S-U data \protect\cite{Abr}, the black triangles represent
the 1996 NA50 Pb-Pb data, the white squares the 1996 analysis with
minimum bias and the black points the 1998 analysis with minimum
bias \protect\cite{Abr50}, but the data are "cleaned out" from the
contribution of $J/\Psi$ scattering in the nuclear matter in
accordance with (\ref{survgas}).} \label{Fig.9.}
\end{minipage}
\hspace{\fill}
\begin{minipage}[t]{75mm}
{\psfig{figure= 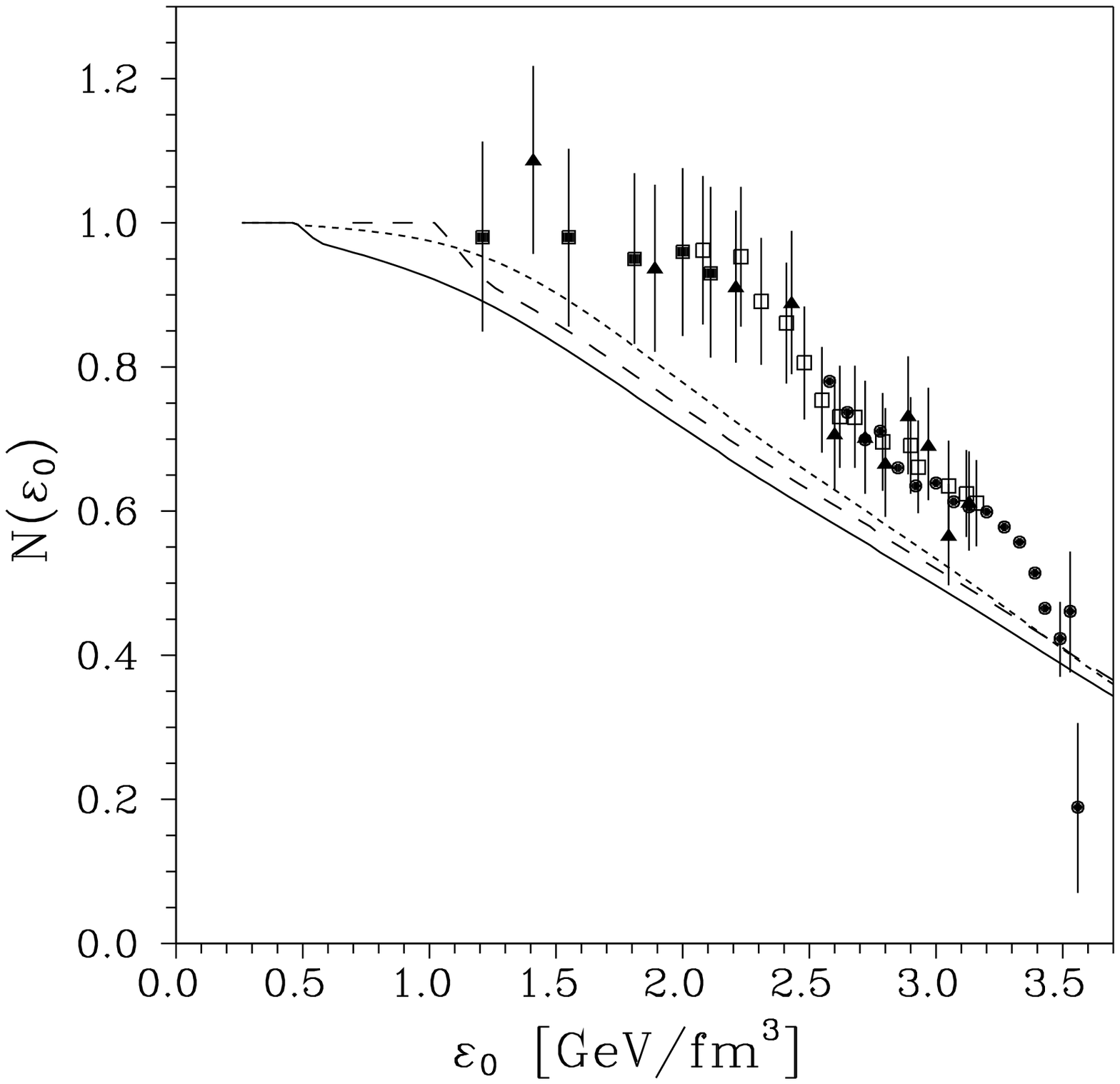 ,height=\textwidth,width=\textwidth}}
 \caption{Same as Fig.\,\ref{Fig.9.} but for $\sigma_{b}=5$ mb and
$\sigma_{m}=3.33$ mb. } \label{Fig.10.}
\end{minipage}
\end{figure}

Now we will include the finite-size effects into our model, i.e.
we will take into account that the realistic hadron gas has a
finite transverse size. This will be done in form of the
rarefaction wave moving inward $S_{eff}$ with the sound velocity
$c_{s}$. How to obtain this velocity has been mentioned in Sec.2
(see also \cite{Prorok:2001ut}). With the finite-size effects
included, the final expression for $J/\Psi$ survival factor ${\cal
N}_{h.g.}(\epsilon_{0})$ will be given by (\ref{15}) . To make our
investigations much more realistic we will also include the
possible $J/\Psi$ disintegration in nuclear matter, which should
increase $J/\Psi$ suppression by about $10 \%$
\cite{Gerschel:1988wn}. But to draw also S-U data in figures,
instead of multiplying ${\cal N}_{h.g.}$ by ${\cal N}_{n.m.}$
given by (\ref{7}), we divide ${\cal N}_{exp}$ by appropriate
${\cal N}_{n.m.}$, i.e. we define "the experimental $J/\Psi$
hadron gas survival factor" as

\begin{equation}
\tilde{{\cal N}}_{exp}= \exp \left\{ \sigma_{J/\psi N} \rho_{0} L
\right\} \cdot {\cal N}_{exp}\;. \label{survgas}
\end{equation}

\noindent and values of this factor are drawn in
Figs.\,\ref{Fig.9.}-\ref{Fig.10.} and
Figs.\,\ref{Fig.12.}-\ref{Fig.13.} as the experimental data.

We shall consider the Woods-Saxon nuclear matter density
distribution \cite{Jager}, here. The results of numerical
estimations of (\ref{15})  and (\ref{survgas}) are depicted in
Figs.\,\ref{Fig.9.}-\ref{Fig.10.} for two values of the
charmonium-baryon cross-section $\sigma_{b}=4,\;5$ mb and the
initial baryon number density $n_{B}^{0}=0.25,\;0.65$ fm$^{-3}$.
The curves for $n_{B}^{0}=0.05$ fm$^{-3}$ almost cover the curves
for $n_{B}^{0}=0.25$ fm$^{-3}$, so for clearness of the figure we
do not draw them. The two values of the speed of sound are the
maximal values of this quantity possible in the range
$[T_{f.o.}=140$ MeV, $T_{0,max}]$ for the above-mentioned two
cases of $n_{B}^{0}$. In fact, we have checked that the results
almost do not depend on $c_{s}$ allowed in the range.

It has turned out also that in the case of the transverse
expansion, the results almost do not depend on the $T_{f.o.}$ (for
$T_{f.o.} \in [100, 140]$ MeV). This is because the freeze-out
time resulting from the transverse expansion, $t_{f.o.,trans} =
R_{A}/c_{s}$ (if we assume a central collision and $c_{s}$
constant), is of the order of the freeze-out time resulting from
the longitudinal expansion for $T_{f.o.}=140$ MeV. Namely, for Pb
and $c_{s}=0.45$ we have $t_{f.o.,trans} \cong 15.8$ fm, which is
very similar to values of $t_{f.o.}$ for $T_{f.o.}=140$ MeV given
earlier. For $T_{f.o.}=100$ MeV, $t_{f.o.} = 111.0,\;101.0,\;72.5$
fm for $n_{B}^{0}= 0.05,\;0.25,\;0.65$ fm$^{-3}$ respectively, so
the hadron gas ceases because of the transverse expansion much
earlier.

\begin{figure}
\begin{center}{
{\epsfig{file=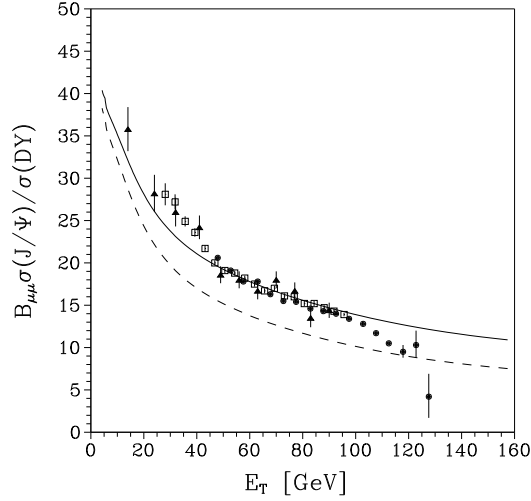,width=7cm}} }\end{center}
\caption{$J/\Psi$ survival factor times
${B_{\mu\mu}\sigma_{J/\psi}^{pp}} \over {\sigma_{DY}^{pp}}$ in the
longitudinally and transversely expanding hadron gas for the
Woods-Saxon nuclear matter density distribution and
$n_{B}^{0}=0.25$ fm$^{-3}$, $T_{f.o.}=140$ MeV, $c_{s}=0.45$ and
$r_{0}=1.2$ fm. The curves correspond to $\sigma_{b}=4$ mb (solid)
and $\sigma_{b}=5$ mb (dashed). The black triangles represent the
1996 NA50 Pb-Pb data, the white squares the 1996 analysis with
minimum bias and the black points the 1998 analysis with minimum
bias \protect\cite{Abr50}. } \label{Fig.11.}
\end{figure}

\begin{figure}[htb]
\begin{minipage}[t]{75mm}
{\psfig{figure= 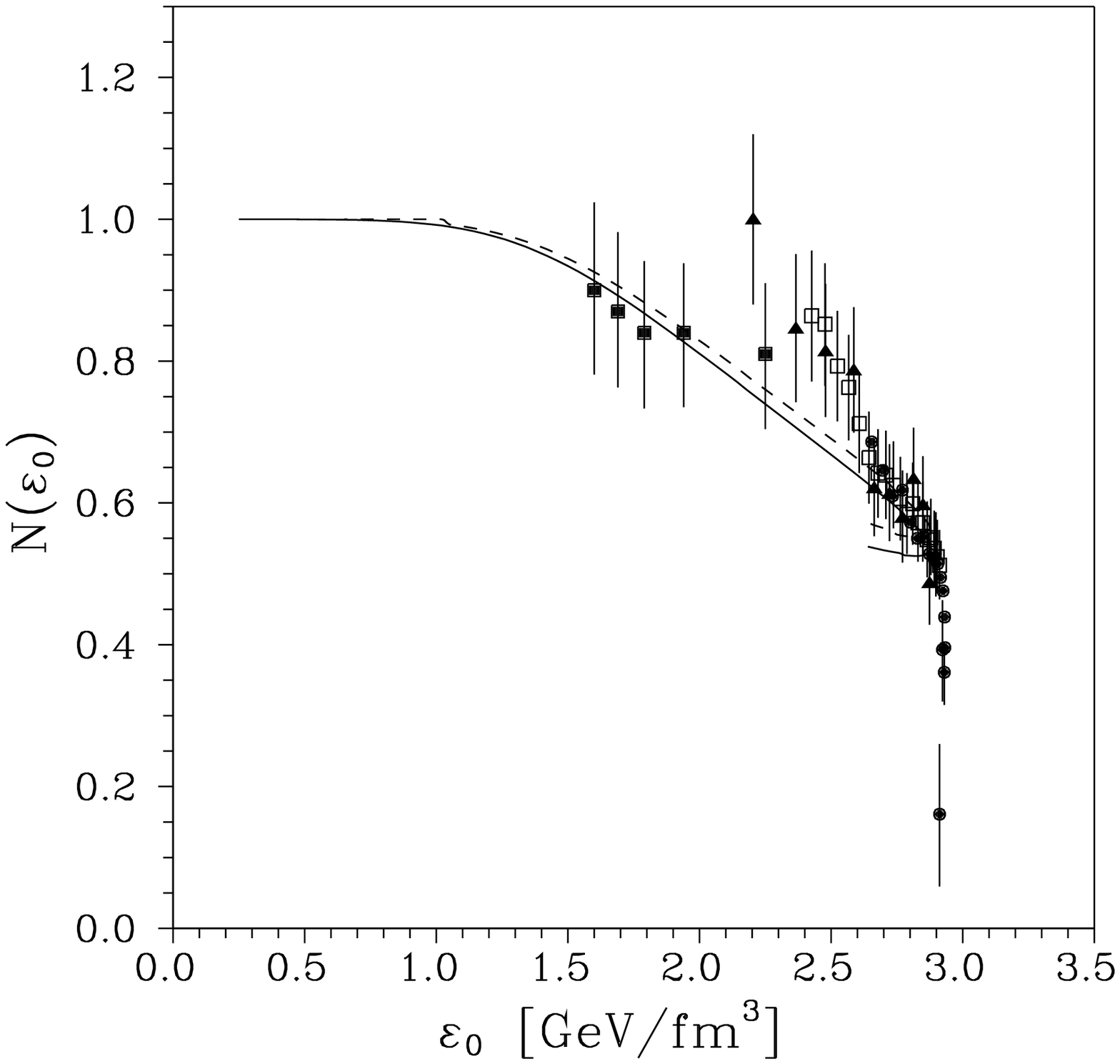,height=\textwidth,width=\textwidth}}
 \caption{$J/\Psi$ suppression in the longitudinally and
transversely expanding hadron gas for the Woods-Saxon nuclear
matter density distribution and $\sigma_{b}=4$mb,
$\sigma_{m}=2.66$ mb, $T_{f.o.}=140$ MeV and $r_{0}=1.2$ fm but
for $\epsilon_{0}(b)$ given by (\ref{etpart}). The curves
correspond to $n_{B}^{0}=0.25$ fm$^{-3}$, $c_{s}=0.45$ (solid) and
$n_{B}^{0}=0.65$ fm$^{-3}$, $c_{s}=0.46$ (dashed). The black
squares represent the NA38 S-U data \protect\cite{Abr}, the black
triangles represent the 1996 NA50 Pb-Pb data, the white squares
the 1996 analysis with minimum bias and the black points the 1998
analysis with minimum bias \protect\cite{Abr50}, but the data are
"cleaned out" from the contribution of $J/\Psi$ scattering in the
nuclear matter in accordance with (\ref{survgas}).}
\label{Fig.12.}
\end{minipage}
\hspace{\fill}
\begin{minipage}[t]{75mm}
{\psfig{figure= 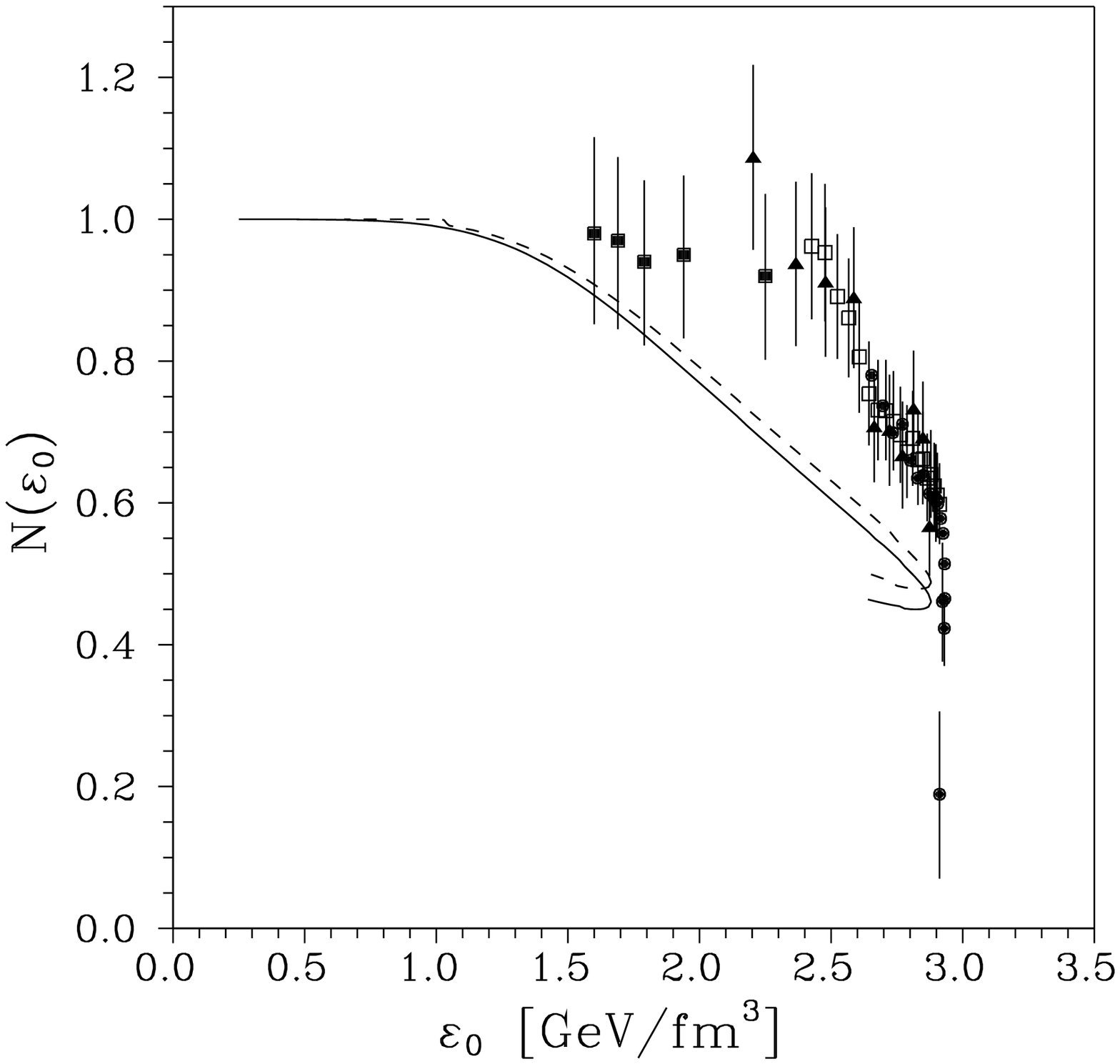 ,height=\textwidth,width=\textwidth}}
 \caption{Same as Fig.\,\ref{Fig.12.} but for $\sigma_{b}=5$ mb and
$\sigma_{m}=3.33$ mb. } \label{Fig.13.}
\end{minipage}
\end{figure}

Generally, taking into account also the transverse expansion
changes the final (theoretical) pattern of $J/\Psi$ suppression
qualitatively. First of all, the curves for the case including the
transverse expansion are not convex, in opposite to the case with
the longitudinal expansion only, where the curves are. But still,
theoretical curves do not fall steep enough at high $\epsilon_{0}$
to cover the data area completely. Nevertheless, from
Figs.\,\ref{Fig.9.}-\ref{Fig.10.} we can see that for some choice
of parameters, namely for $\sigma_{b}$ somewhere between 4 and 5
mb and for $r_{0}=1.12$ fm, we would obtain a quite satisfactory
curve. And we should remember that since we have one overall
charmonium-baryon cross-section $\sigma_{b}$, our final results
underestimate the suppression (for $\chi-,\;\psi'-baryon$
scattering the cross-section should be greater than for $J/\Psi$).
To support our conclusion in more visible way we present main
results from Figs.\,\ref{Fig.9.}-\ref{Fig.10.} in
Fig.\,\ref{Fig.11.}, where original data \cite{Abr50} for
${B_{\mu\mu}\sigma_{J/\psi}^{PbPb}} \over {\sigma_{DY}^{PbPb}}$
and $J/\Psi$ survival factors given by (\ref{6}) multiplied by
${B_{\mu\mu}\sigma_{J/\psi}^{pp}} \over {\sigma_{DY}^{pp}}$ and
now as functions of $E_{T}$ are presented. The change of the
variable from $\epsilon_{0}$ to $E_{T}$ has been done with the use
of (\ref{etwound}) and $\epsilon_{0}=\epsilon_{0}(b)$ expressed by
the solid line in Fig.\,\ref{Fig.4.}.

Note  that the main disagreement with the data reveals in the last
experimental point of the 1998 analysis \cite{Abr50} (see
Figs.\,\ref{Fig.5.}-\ref{Fig.11.}). But the error bar of this
point is very wide. Additionally, there is some contradiction in
positions of the last three points of the 1998 data. In fact, the
middle point goes up above the left one and the last falls far
below the two earlier. This suggests that more experimental data
is needed in the region of high $\epsilon_{0}$ to state definitely
whether the abrupt fall of the experimental suppression factor
takes place or not there.

Now we repeat again the numerical estimations of (\ref{15}) for
the case with the finite-size effects and the Woods-Saxon nuclear
matter density distribution included, but for the dependence of
$\epsilon_{0}$ on $b$ given by (\ref{etpart}). The results are
presented in Figs.\,\ref{Fig.12.}-\ref{Fig.13.}. Note that the
theoretical curves are two-valued around $\epsilon_{0}= 2.8$
GeV/fm$^{3}$. This is the result of our approximation of
$\epsilon_{0}(b)$ by (\ref{etpart}). This expression allows for
two different values of $b$, which give the same $\epsilon_{0}$ in
some range of the impact parameter $b$. This is shown in
Fig.\,\ref{Fig.4.} (short-dashed line). We can see that for $b
\leq 7.9$ fm there are two different values $b_{1}$ and $b_{2}$
such that $\epsilon_{0}(b_{1}) = \epsilon_{0}(b_{2})$. This causes
that results plotted in Figs.\,\ref{Fig.12.}-\ref{Fig.13.} are
qualitatively different from those presented in
Figs.\,\ref{Fig.9.}-\ref{Fig.10.} in the region close to the
maximal $\epsilon_{0}$ reached at the collision. Comparing
Figs.\,\ref{Fig.9.}-\ref{Fig.10.} with
Figs.\,\ref{Fig.12.}-\ref{Fig.13.}, we can see also that the
pattern of $J/\Psi$ suppression depends on the shape of
$\epsilon_{0}$ as a function of $b$. Clarification of this
dependence would be very helpful to obtain more realistic picture
of the $J/\Psi$ dissociation in hadron medium during the heavy-ion
collisions.

As a final remark, we think that it is difficult to exclude
$J/\Psi$ scattering in the hot hadron gas entirely, as the reason
for the observed $J/\Psi$ suppression at this point (see also
\cite{Capella:2000zp}). In our model the most crucial parameter is
the $charmonium-baryon$ inelastic cross-section and the final
results depend on its value substantially. Therefore it is of the
greatest importance to establish how this cross-section behaves in
the hot hadron environment. Some work has been done into this
direction \cite{Kharzeev:1994pz,Martins:1995hd,Matinian:1998cb},
but results presented there differ from each other and are based
on different models. However, the newest estimations of
$\pi+J/\Psi$, $\rho+J/\Psi$ and $J/\Psi+N$ cross-sections at high
invariant collision energies
\cite{Tsushima:2000cp,Sibirtsev:2000aw} agree with the values of
$\sigma_{b}$ and $\sigma_{m}$ assumed in our model. We would like
to add also at this point that the $charmonium-hadron$ inelastic
cross-sections have been considered as constant quantities here.
For sure, they should not be constant and the results of just
mentioned papers \cite{Tsushima:2000cp,Sibirtsev:2000aw} suggests
that they are not, indeed. They are growing functions of the
invariant collision energy $\sqrt{s}$. So, the naive reasoning
should direct us to the conclusion that the increase of
$\epsilon_{0}$ (or in other words $E_{T}$) causes the increase of
the invariant collision energy $\sqrt{s}$ on the average and
further the increase of the $charmonium-hadron$ inelastic
cross-sections. This could influence the final patterns of
$J/\Psi$ suppression in such a way that $J/\Psi$ survival factor
would behave according to the solid curve of Fig.\,\ref{Fig.11.}
for low $\epsilon_{0}$ ($E_{T}$) but then, as the
$charmonium-hadron$ inelastic cross-sections would increase with
$\epsilon_{0}$ ($E_{T}$), the factor would go closer to the dashed
curve of Fig.\,\ref{Fig.11.} for high $\epsilon_{0}$ ($E_{T}$).
So, the experimental pattern of $J/\Psi$ suppression could be
recovered in this way. Therefore, as a final conclusion we can say
that it is difficult to ruled out the conventional explanations of
$J/\Psi$ suppression completely, at present.

We would like to stress again that the behaviour of the
experimental $J/\Psi$ suppression factor at high $E_{T}$ (or
otherwise at high $\epsilon_{0}$) has not been clear yet. In fact,
the abrupt fall of this factor (what could suggest the appearance
of the quark-gluon plasma) is indicated only by the one point (the
last) of the 1998 NA50 analysis \cite{Abr50}. So, to draw a
definite conclusion more experimental data far above $E_{T}=120$
GeV are needed. This region will be reached in upcoming RHIC runs
and their results should answer the question: is the $J/\Psi$
suppression a signature of the existence of the quark-gluon
plasma, or not?.

{\it Note added.} When our paper was completed we became aware of
\cite{Wong:2001if} where the twin figure (denoted as Fig.5 there)
to our Fig.\,\ref{Fig.11.} was presented. But the appearance of
the quark-gluon plasma is the main reason for $J/\Psi$ suppression
there. It is also claimed that results shown in that figure
"provide evidence for the production of the quark-gluon plasma in
central high-energy Pb-Pb collisions". This entirely confirms our
conclusion that the status of $J/\Psi$ suppression as a signal for
the quark-gluon plasma appearance is far from being clear at
present.

\acknowledgments{We would like to thank Dr K. Redlich for very
helpful discussions.

Work supported in part by the Polish Committee for Scientific
Research under contract KBN-2~P03B~030~18\,.}

\end{document}